\documentstyle[12pt]{article}

\input{tcilatex}

\begin{document}

\begin{titlepage}
\vspace{1.5cm} 
\begin{center}
{\LARGE\bf  Knizhnik-Zamolodchikov-Bernard equations connected with 
the eight-vertex model}
\vspace{1cm}

\begin{center}
{\large\bf H. Babujian}$^\dag$\footnote{babujian@Lx2.YerPHI.AM}
\hspace{.4cm}
{\large\bf A. Lima-Santos}$^\ddag$\footnote{dals@power.ufscar.br}

and

\hspace{.2cm}
{\large\bf R. H. Poghossian}$^\dag$\footnote{poghos@moon.YerPHI.AM}
\end{center}

\vspace{1cm}

\begin{center}
{\bf $\vphantom{t}^\dag$Yerevan Physics Institute, Alikhanian Brothers 2, 
Yerevan, 375036 Armenia}
\end{center}

\begin{center}
{\bf $\vphantom{t}^\ddag$Departamento de F\'{\i}sica Universidade Federal de 
S\~ao Carlos \\ 13569-905 S\~ao Carlos, Brasil}
\end{center}

\vspace{2.0cm}
\begin{abstract}
Using quasiclassical limit of Baxter's 8 - vertex  R - matrix, an 
elliptic generalization of the Knizhnik-Zamolodchikov equation is 
constructed. Via Off-Shell Bethe ansatz an integrable representation 
for this equation is obtained. It is shown that there exists a 
gauge transformation connecting this equation with 
Knizhnik-Zamolodchikov-Bernard equation for SU(2)-WZNW model on torus.
 \end{abstract}
 \end{center}
\end{titlepage}

\newpage 

\section{Introduction}

Conformal Field Theory (CFT) describes the critical behavior of
two-dimensional statistical systems, many of which are integrable. Usually
correlation functions are governed by linear differential equations which
are derived from degeneracy relations of the underlying symmetry algebra
(see e.g.\cite{BPZ}, \cite{KZ}, \cite{ZP}, \cite{P}). In integrable models
of statistical mechanics \cite{Baxter}, the main object is the ${\cal R}$%
-matrix which depends on a spectral parameter $\lambda $ and acts on tensor
product $V\otimes V$ for some vector space $V$. The main condition on ${\cal %
R}$, leading to integrability is the Yang-Baxter equation 
\begin{equation}
{\cal R}_{12}(\lambda ){\cal R}_{13}(\lambda +\mu ){\cal R}_{23}(\mu )= 
{\cal R}_{23}(\mu ){\cal R}_{13}(\lambda +\mu ){\cal R}_{12}(\lambda )
\label{eq1.1}
\end{equation}
It is assumed that (\ref{eq1.1}) is defined on threefold tensor product $%
V^{1}\otimes V^{2}\otimes V^{3}$. Here $V^{1}$, $V^{2}$, $V^{3}$ are some
copies of $V$ and lower indices of ${\cal R}$ show the components of tensor
product on which ${\cal R}$-matrix acts non trivially. The Yang-Baxter
equation implies the commutativity of transfer matrices constructed out of $%
{\cal R}$ for two arbitrarily chosen values of spectral parameter $\lambda$.
If ${\cal R}$ has extra dependence on some (Plank-like) parameter $\eta $ so
that ${\cal R}=1+\eta \ r+{\rm o}(\eta ^{2})$, as $\eta \rightarrow 0$, then
the classical $r$-matrix obeys the classical Yang-Baxter equation 
\begin{equation}
\lbrack r_{12}(\lambda ),r_{13}(\lambda +\mu )+r_{23}(\mu )]+[r_{13}(\lambda
+\mu ),r_{23}(\mu )]=0  \label{eq1.2}
\end{equation}
This equation has the following relation with conformal field theory. In the
skew-symmetric case $r_{21}(-\lambda )=-r_{12}(\lambda )$, it is the
compatibility condition for the system of linear differential equations 
\begin{equation}
\frac{d\Psi }{d\lambda _{i}}=\sum_{j\neq i}r_{ij}(\lambda _{i}-\lambda
_{j})\Psi  \label{eq1.3}
\end{equation}
for a function $\Psi (\lambda _{1},\lambda _{2},\cdots ,\lambda _{N})$ with
values in $V^{1}\otimes \cdots \otimes V^{N}$. In the rational case \cite{BD}
, very simple skew-symmetric solutions are known: $r(\lambda )=C/\lambda $,
where $C\in {\rm g}\otimes {\rm g}$ is a symmetric invariant tensor of a
finite dimensional Lie algebra ${\rm g}$ acting on a representation space $V$%
. Then the corresponding system of differential equations (\ref{eq1.3}) is
the Knizhnik-Zamolodchikov (KZ) equation for the conformal blocks of the
Wess-Zumino-Novikov-Witten (WZNW) model on the sphere \cite{KZ}. Usually the
exactly integrable homogeneous vertex model and its connection with the
conformal field theory \cite{deVega} or quantum field theory \cite{Luther}
has been considered. Here we consider the inhomogeneous vertex model, where
to each vertex we associate two parameters: the global spectral parameter $%
\lambda $ and the disorder parameter $z$. The vertex weight matrix ${\cal R}$
depends on $\lambda -z$. Hence the transfer matrix of the vertex model now
depends on the disorder parameters $z_{i}$ , $i=1,\ldots ,N$.

If we have some rational solution of Yang-Baxter equation (\ref{eq1.1}) and
the transfer matrix $T(\lambda |z)$, then by construction of the algebraic
Bethe ansatz \cite{FT} we have an equation 
\begin{equation}
T(\lambda |z)\Phi (\lambda _{1,\cdots ,}\lambda _{n}|z)=\Lambda (\lambda
,\lambda _{1},\cdots ,\lambda _{n}|z)\Phi (\lambda _{1},\cdots ,\lambda
_{n}|z)-\sum_{\alpha =1}^{n}\frac{F_{\alpha }\Phi ^{\alpha }}{\lambda
-\lambda _{\alpha }}  \label{eq1.4}
\end{equation}
where 
\[
\Phi (\lambda _{1},\cdots ,\lambda _{n}|z)=\Phi (\lambda _{1},\cdots
,\lambda _{n}|z_{1},\cdots ,z_{N}) 
\]
is the Bethe wave vector, 
\[
\Phi ^{\alpha }=\Phi (\lambda _{1},\cdots ,\lambda _{\alpha -1},\lambda
,\lambda _{\alpha -1},\cdots \lambda _{n}|z_{1},\cdots ,z_{N}) 
\]
and $F_{\alpha }(\lambda _{1},\cdots ,\lambda _{n}|z_{1},\cdots ,z_{N})$, $%
\Lambda (\lambda ,\lambda _{1},\cdots ,\lambda _{n}|z_{1},\cdots ,z_{N})$
are some c-valued functions. In the Bethe ansatz we impose the condition $%
F_{\alpha }=0$. Under this condition the wave vector $\Phi $ becomes an
eigenvector and $\Lambda $ an eigenvalue of transfer matrix $T(\lambda |z)$.
In general case, when the condition $F_{\alpha}=0$ doesn't imposed,
following \cite{B0}-\cite{B3} we'll refer to (\ref{eq1.4}) as Off-Shell
Bethe Ansatz Equation (OSBAE).Using OSBAE (\ref{eq1.4}) in quasi classical
limit $\eta \rightarrow 0$, in \cite{B0}-\cite{B3} the solution of the
Knizhnik-Zamolodchikov equation for the $N$-point correlation function $\Psi
(z_{1},\ldots ,z_{N})$ in WZNW theory \cite{KZ} 
\begin{equation}
\kappa \frac{\partial \Psi }{\partial z_{k}}=\sum_{i\neq k}^{N} \frac{%
t_{k}^{a}\otimes t_{i}^{a}}{z_{k}-z_{i}} \Psi  \label{eq1.5}
\end{equation}
has been constructed. In quasi classical limit (\ref{eq1.4}) transforms into
OSBAE for the Gaudin non-local hamiltonians \cite{Gaudin} $H_{k}$ , $%
k=1,\ldots ,N$: 
\begin{equation}
H_{k}=\sum_{i\neq k}^{N}\frac{t_{k}^{a}\otimes t_{i}^{a}}{z_{k}-z_{i}}
\label{eq1.6}
\end{equation}
where as in (\ref{eq1.5}) $t_{i}^{a}$, $a=1,...,\dim ({\rm g})$, represent
the generators of the simple Lie algebra ${\rm g}$ and act non-trivially in
the representation spaces $V^{i}$, $i=1,...,N$. The vector-valued function $%
\Psi(z_{1},\ldots ,z_{N})$ is the holomorphic part (or conformal block) of
the $N$-point correlation function in WZNW theory.

The solution of (\ref{eq1.5}) has the form 
\begin{equation}
\Psi (z_{1},\ldots ,z_{N})=\oint \cdots \oint {\cal X}(\lambda
_{1},...,\lambda _{n}|z)\phi (\lambda _{1},...,\lambda _{n}|z)d\lambda
_{1}\cdots d\lambda _{n}  \label{eq1.7}
\end{equation}
where $\phi (\lambda _{1},...,\lambda _{n}|z)$ is the semiclassical limit of
the Bethe vector $\Phi (\lambda _{1},...,\lambda _{n}|z)$ and in fact it is
the Bethe wave vector for Gaudin magnets (\ref{eq1.6}), but off mass shell.
The scalar function ${\cal X}(\lambda _{1},...,\lambda _{n}|z)$ is
constructed from the semiclassical limit of the $\Lambda (\lambda
=z_{k};\lambda _{1},...,\lambda _{n}|z)$ and $F_{\alpha}$ (for more details
see \cite{B0}-\cite{B3}).

This representation of the N-point correlation function in WZNW theory shows
an intriguing connection between the inhomogeneous vertex models and the
WZNW theory. A deeper understanding of this fine structure may provide us
with more complete knowledge about exact integrability and conformal field
theory in two-dimensions.

In this paper we consider OSBAE for the inhomogeneous eight-vertex model and
corresponding $XYZ$ Gaudin magnet, which provide the solution of an elliptic
generalization of the Knizhnik-Zamolodchikov equation 
\begin{eqnarray}
\kappa (\frac{\partial }{\partial z_{i}}-\Gamma_{i})\Psi= \sum_{j\neq
i}r_{ij}(z_i-z_j)\Psi-\frac{\partial}{\partial w} (\Sigma_{j}\Psi),
\label{eq1.8}
\end{eqnarray}
where $r_{ij}(\lambda)$ is the elliptic quasi classical r-matrix. $\Sigma_i$%
, $\Gamma_i$ are some operators acting non trivially only on i-th component
of N-fold tensor product ${\bf C}^2\otimes \cdots {\bf C}^2$, where the
vector $\Psi$ takes its values. The exact form of $\Sigma_i$, $\Gamma_i$
will be specified in section $6$.

The distinguishing feature of elliptic case is that the vector $\Psi$
depends not only on $z_1 ,\ldots ,z_N$ as in trigonometric and rational
cases, but also on elliptic moduli and on some auxiliary parameter $w$, the
meaning of which will be clarified in the main text.

In Section $2$, we present the BAE and OSBAE for the inhomogeneous
eight-vertex model and its quasi classical limit spin-$1/2$ XYZ Gaudin
model. In Section $3$, we construct EKZ equation (\ref{eq1.8}) and its
solution. In Section $4$, we discuss the connection of EKZ equation and the
Knizhnik-Zamolodchikov-Bernard equation for $su(2)$ WZNW theory on torus and
finally we left the last section to some remarks and conclusions .

\section{Algebraic Bethe ansatz for the inhomogeneous eight-vertex model}

In this section we modify Algebraic Bethe Ansatz (ABE) technique for $8$%
-vertex model developed in \cite{FT} in order to adjust it to the
inhomogeneous case.

The inhomogeneous eight-vertex model is parameterized by an anisotropy
parameter $\eta $, elliptic modulus $\tau $ and the shifts of the spectral
parameter (inhomogeneity parameters $z_{i}$).

The algebraic Bethe ansatz solution of the inhomogeneous eight-vertex model
is closely related to the homogeneous case. As we'll be clear from the
further consideration, the only modification one have to do, is local
shifting of the spectral parameter $\lambda \rightarrow \lambda -z_i$ and
auxiliary parameters $s,t\rightarrow s-z_i,t+z_i$. Note one nonessential
difference from the conventions of \cite{FT}, that we label the sites in a
horizontal row of lattice from left to right (as in \cite{Baxter1}). We
define a local transition matrix ${\cal L}_i(\lambda -z_i)$ by 
\begin{equation}
{\cal L}_i(\lambda -z_i)=\sum_{a=1}^4w_a\ \sigma ^a\otimes \sigma
_i^a=\left( 
\begin{array}{ll}
w_4\sigma _i^4+w_3\sigma _i^3 & w_1\sigma _i^1-iw_2\sigma _i^2 \\ 
w_1\sigma _i^1+iw_2\sigma _i^2 & w_4\sigma _i^4-w_3\sigma _i^3
\end{array}
\right)   \label{eq2.1}
\end{equation}
where the $\sigma ^i$ are the spin-$1/2$ Pauli matrices, which in the basis 
\begin{equation}
{\rm e}^{+}=\left( 
\begin{array}{l}
1 \\ 
0
\end{array}
\right) ,\qquad {\rm e}^{-}=\left( 
\begin{array}{l}
0 \\ 
1
\end{array}
\right)   \label{eq2.2}
\end{equation}
have the standard form 
\begin{equation}
\sigma ^1=\left( 
\begin{array}{ll}
0 & 1 \\ 
1 & 0
\end{array}
\right) ,\quad \sigma ^2=\left( 
\begin{array}{lr}
0 & -i \\ 
i & 0
\end{array}
\right) ,\quad \sigma ^3=\left( 
\begin{array}{lr}
1 & 0 \\ 
0 & -1
\end{array}
\right)   \label{eq2.3}
\end{equation}
and $\sigma ^4=I$ it is the $2\times 2$ identity matrix. For the
inhomogeneous case the coefficients are parameterized as 
\begin{eqnarray}
w_4+w_3 &=&\rho (\lambda -z_i)\Theta (2\eta )\ \Theta (\lambda -z_i-\eta )\
H(\lambda -z_i+\eta )  \nonumber \\
w_4-w_3 &=&\rho (\lambda -z_i)\Theta (2\eta )\ H(\lambda -z_i-\eta )\ \Theta
(\lambda -z_i+\eta )  \nonumber \\
w_1+w_2 &=&\rho (\lambda -z_i)H(2\eta )\ \Theta (\lambda -z_i-\eta )\ \Theta
(\lambda -z_i+\eta )  \nonumber \\
w_1-w_2 &=&\rho (\lambda -z_i)H(2\eta )\ H(\lambda -z_i-\eta )\ H(\lambda
-z_i+\eta )  \label{eq2.4}
\end{eqnarray}
where $\Theta (\lambda )$ and $H(\lambda )$ are the Jacobi theta-functions, 
\begin{equation}
g(\lambda )=H(\lambda )\Theta (\lambda )  \label{eq2.5}
\end{equation}
and we have introduced the normalizing function 
\begin{equation}
\rho (\lambda )=\frac{g(K)}{\Theta (2\eta )g(K+\eta )g(\lambda )}
\label{eq2.6}
\end{equation}
in order to fix Boltzmann weight $w_4\equiv 1$ ($K$ is the half-period of
Jacobi function $\Theta (\lambda ))$.

The matrix ${\cal L}_{i}(\lambda -z_{i})$ satisfies the Yang-Baxter relation 
\begin{equation}
{\cal R}(\lambda ,\mu ){\cal L}_{i}(\lambda -z_{i})\otimes {\cal L}_{i}(\mu
-z_{i})= {\cal L}_{i}(\mu -z_{i})\otimes {\cal L}_{i}(\lambda -z_{i}){\cal R}%
(\lambda ,\mu)  \label{eq2.7}
\end{equation}
where the ${\cal R}$- matrix has the form 
\begin{equation}
{\cal R}(\lambda ,\mu )=\left( 
\begin{array}{llll}
a & 0 & 0 & d \\ 
0 & b & c & 0 \\ 
0 & c & b & 0 \\ 
d & 0 & 0 & a
\end{array}
\right)  \label{eq2.8}
\end{equation}
with 
\begin{eqnarray}
a(\lambda ,\mu ) &=&\Theta (2\eta )\ \Theta (\lambda -\mu -\eta )\ H(\lambda
-\mu +\eta )  \nonumber \\
b(\lambda ,\mu ) &=&\Theta (2\eta )\ H(\lambda -\mu -\eta )\ \Theta (\lambda
-\mu +\eta )  \nonumber \\
c(\lambda ,\mu ) &=&H(2\eta )\ \Theta (\lambda -\mu -\eta )\ \Theta (\lambda
-\mu +\eta )  \nonumber \\
d(\lambda ,\mu ) &=&H(2\eta )\ H(\lambda -\mu -\eta )\ H(\lambda -\mu +\eta )
\label{eq2.9}
\end{eqnarray}

The operator matrix 
\begin{eqnarray}
{\cal T}_{N}(\lambda |z) &=&{\cal L}_{1}(\lambda -z_{1}) {\cal L}
_{2}(\lambda -z_{2})\cdots{\cal L}_{N}(\lambda -z_{N})  \nonumber \\
&=&\left( 
\begin{array}{ll}
A_{N}(\lambda |z) & B_{N}(\lambda |z) \\ 
C_{N}(\lambda |z) & D_{N}(\lambda |z)
\end{array}
\right)  \label{eq2.10}
\end{eqnarray}
is called the monodromy matrix and the operator 
\begin{equation}
T_{N}(\lambda |z)=A_{N}(\lambda |z)+D_{N}(\lambda |z)  \label{eq2.11}
\end{equation}
is called the transfer matrix. In this text we will use a compact notation
for the arguments with the shifted spectral parameter: $(\lambda
|z)=(\lambda -z_{1},\cdots ,\lambda -z_{N})$.

In contrast to the trigonometric and rational cases, the eight-vertex model
does not have a unique vacuum vector in its local state space. We follow 
\cite{FT} to overcome this difficulty by introducing a family of gauge
transformations with parameters $s$ and $t$ as follows:

\begin{eqnarray}
{\cal L}_{i}(\lambda -z_{i}) &\rightarrow &{\cal L}_{i}^{l}(\lambda
,z_{i};s,t) =  \nonumber \\
&=&M_{l+i-1}^{-1}(\lambda;s,t){\cal L}_{i}(\lambda
-z_{i})M_{l+i}(\lambda;s,t)  \nonumber \\
&=&\left( 
\begin{array}{ll}
\alpha _{i}^{l}(\lambda ,z_{i};s,t) & \beta _{i}^{l}(\lambda ,z_{i};s,t) \\ 
\gamma _{i}^{l}(\lambda ,z_{i};s,t) & \delta _{i}^{l}(\lambda ,z_{i};s,t)
\end{array}
\right)  \label{eq2.12}
\end{eqnarray}
matrix ${\cal L}_{i}^{l}$ has a local vacuum, $\omega _{i}^{l}$ ,
independent of $\lambda $ that is annihilated for all $\lambda $ by its
lower left element. Again, in contrast to the case of the six-vertex model,
the local vacuum $\omega _{i}^{l}$ is not an eigenvector of $\alpha _{i}^{l}$
and $\delta _{i}^{l}.$ So, the matrices $M_{k}$ must also be chosen in such
a way, that the diagonal elements of ${\cal L}_{i}^{l}$ act on the local
vacuum in the simplest possible way. Following \cite{Baxter}, \cite{FT} we
choose 
\begin{equation}
M_{k}(\lambda ;s,t)=\left( 
\begin{array}{ll}
H(s+2k\eta +\lambda ) & \frac{1}{g(w_k)}H(t+2k\eta -\lambda) \\ 
\Theta (s+2k\eta +\lambda ) & \frac{1}{g(w_k)}\Theta (t+2k\eta-\lambda ),
\end{array}
\right)  \label{eq2.13}
\end{equation}
where $w _{k}=\frac{s+t}{2}+2k\eta -K$. Corresponding local vacuum 
\begin{equation}
\omega _{i}^{l}=H(s+2(l+i)\eta -\eta +z_{i}){\rm e}_{i}^{+}+\Theta
(s+2(l+i)\eta -\eta +z_{i}){\rm e}_{i}^{-}  \label{eq2.14}
\end{equation}
has the properties: 
\begin{eqnarray}
\alpha _{i}^{l}(\lambda, z_{i};s,t)\omega _{i}^{l} &=&
\rho(\lambda-z_{i})\Theta(0) g(\lambda -z_{i}+\eta)\omega _{i}^{l+1} 
\nonumber \\
\delta _{i}^{l}(\lambda, z_{i};s,t)\omega _{i}^{l} &=& \Theta (0)g(\lambda
-z_{i}-\eta)\omega _{i}^{l-1}  \nonumber \\
\gamma _{i}^{l}(\lambda, z_{i};s,t)\omega _{i}^{l} &=&0  \label{eq2.15}
\end{eqnarray}

Now let us define the set of gauge transformed monodromy matrices 
\begin{eqnarray}
{\cal T}_{N}^{l}(\lambda |z;s,t) &=&{\cal L}_{1}^{l}(\lambda ,z_{1};s,t)%
{\cal L}_{2}^{l}(\lambda, z_{2};s,t)\cdots {\cal L} _{N}^{l}(\lambda,
z_{N};s,t)  \nonumber \\
&=&\left( 
\begin{array}{ll}
A_{N}^{l}(\lambda |z;s,t) & B_{N}^{l}(\lambda |z;s,t) \\ 
C_{N}^{l}(\lambda |z;s,t) & D_{N}^{l}(\lambda |z;s,t)
\end{array}
\right)  \label{eq2.16}
\end{eqnarray}
>From the local formulae (\ref{eq2.13}-\ref{eq2.15}) one can derive that
matrix elements of (\ref{eq2.16}) satisfy the relations 
\begin{eqnarray}
A_{N}^{l}(\lambda |z;s,t)\Omega _{N}^{l} &=&\prod_{i=1}^{N} \left[
\rho(\lambda-z_{i})\Theta (0) g(\lambda-z_{i}+\eta )\right] \Omega _{N}^{l+1}
\nonumber \\
D_{N}^{l}(\lambda |z;s,t)\Omega _{N}^{l} &=&
\prod_{i=1}^{N}\left[\rho(\lambda-z_{i}) \Theta (0)g(\lambda-z_{i}-\eta
)\right]\Omega _{N}^{l-1}  \nonumber \\
C_{N}^{l}(\lambda |z;s,t)\Omega _{N}^{l} &=&0  \label{eq2.17}
\end{eqnarray}
Where the set of generating vectors $\left\{ \Omega _{N}^{l}\right\}
_{l=-\infty }^{l=\infty }$ is defined by 
\begin{equation}
\Omega _{N}^{l}=\omega _{1}^{l}\otimes \omega _{2}^{l}\otimes \cdots \otimes
\omega _{N}^{l}  \label{eq2.18}
\end{equation}
It is useful to introduce a collection of generalized monodromy matrices 
\begin{eqnarray}
{\cal T}_{k,l}(\lambda |z;s,t) &=&M_{k}^{-1}(\lambda |z;s,t){\cal T}_{N}
(\lambda |z)M_{l}(\lambda |z;s,t)  \nonumber \\
&=&\left( 
\begin{array}{ll}
A_{k,l}(\lambda |z;s,t) & B_{k,l}(\lambda |z;s,t) \\ 
C_{k,l}(\lambda |z;s,t) & D_{k,l}(\lambda |z;s,t)
\end{array}
\right)  \label{eq2.19} \\
k,l &=&-\infty ,\cdots ,\infty  \nonumber
\end{eqnarray}
The monodromy matrix ${\cal T}_{N}^{l}(\lambda |z;s,t)$ can be written in
the new notation as ${\cal T}_{N+l,l}(\lambda |z;s,t)$ and for all values of 
$l$ 
\begin{equation}
T(\lambda |z)=A_{N}(\lambda |z)+D_{N}(\lambda |z)=A_{l,l}(\lambda
|z;s,t)+D_{l,l}(\lambda |z;s,t),  \label{eq2.20}
\end{equation}

It turns out from the relation 
\begin{equation}
{\cal R}(\lambda ,\mu ){\cal T}_{N}(\lambda |z)\otimes {\cal T}_{N}(\mu |z)= 
{\cal T}_{N}(\mu |z)\otimes {\cal T}_{N}(\lambda |z){\cal R}(\lambda ,\mu )
\label{eq2.21}
\end{equation}
that the permutation relations for $A_{N},B_{N},C_{N}$ and $D_{N}$ , as in
the homogeneous case, lead to simple relations for $A_{k,l},B_{k,l},C_{k,l}.$
and $D_{k,l}$: 
\begin{equation}
B_{k,l-1}(\lambda |z;s,t)B_{k-1,l}(\mu |z;s,t)=B_{k,l-1}(\mu
|z;s,t)B_{k-1,l}(\lambda |z;s,t),  \label{eq2.22}
\end{equation}
\[
A_{k,l}(\lambda |z;s,t)B_{k-1,l+1}(\mu |z;s,t)=\alpha (\lambda ,\mu
)B_{k,l+2}(\mu |z;s,t)A_{k-1,l+1}(\lambda |z;s,t) 
\]
\begin{equation}
+\beta _{l+1}(\lambda ,\mu )B_{k,l+2}(\lambda |z;s,t)A_{k-1,l+1}(\mu |z;s,t)
\label{eq2.23}
\end{equation}
\[
D_{k,l}(\lambda |z;s,t)B_{k-1,l+1}(\mu |z;s,t)=\alpha (\mu ,\lambda)
B_{k-2,l}(\mu |z;s,t)D_{k-1,l+1}(\lambda |z;s,t) 
\]
\begin{equation}
-\beta _{k-1}(\lambda ,\mu )B_{k-2,l}(\lambda |z;s,t)D_{k-1,l+1}(\mu |z;s,t)
\label{eq2.24}
\end{equation}
where 
\begin{equation}
\alpha(\lambda ,\mu )=\frac{g(\lambda -\mu -2\eta )}{g(\lambda -\mu )}
,\qquad \beta _{l}(\lambda ,\mu )=\frac{g(2\eta )g(w _{l}+\lambda-\mu )} {%
g(\lambda -\mu)g(w _{l})}  \label{eq2.25}
\end{equation}

Next, let us consider the Bethe vectors 
\begin{equation}
\Psi _{l}(\lambda _{1},\lambda _{2},\cdots ,\lambda
_{n})=B_{l-1,l+1}(\lambda _{1})\cdots B_{l-n,l+n}(\lambda _{n})\Omega
_{N}^{l-n}  \label{eq2.26}
\end{equation}
where $n=N/2$ ($N$ even).

The Bethe vectors (\ref{eq2.26}) are not eigenstates of the transfer matrix (%
\ref{eq2.20}), but satisfy the relations 
\[
T(\lambda |z)\Psi _{l}(\lambda _{1},\cdots ,\lambda _{n})= 
\]
\[
=\Lambda _{1}(\lambda |z;\lambda _{1},\cdots ,\lambda _{n})\Psi
_{l+1}(\lambda _{1},\cdots,\lambda _{n} )+ \Lambda _{2}(\lambda |z;\lambda
_{1},\cdots ,\lambda _{n})\Psi _{l-1}(\lambda _{1},\cdots,\lambda _{n} ) 
\]
\begin{equation}
+\sum_{\alpha=1 }^{n }\left[ \Lambda _{1,l}^{\alpha}(\lambda |z;\lambda
_{1},\cdots ,\lambda _{n})\Psi _{l+1}^{\alpha}+\Lambda
_{2,l}^{\alpha}(\lambda |z;\lambda _{1}, \cdots ,\lambda_{n})\Psi
_{l-1}^{\alpha}\right]  \label{eq2.27}
\end{equation}
where 
\begin{eqnarray}
\Lambda _{1}(\lambda |z;\lambda _{1},\cdots ,\lambda _{n})
=\prod_{i=1}^{N}\left[\rho(\lambda-z_{i})\Theta (0)g(\lambda -z_{i}+ \eta
)\right] \prod_{\alpha =1}^{n}\alpha(\lambda ,\lambda _{\alpha }) \quad \quad
\nonumber \\
\Lambda _{2}(\lambda |z;\lambda _{1},\cdots ,\lambda _{n})
=\prod_{i=1}^{N}\left[\rho (\lambda-z_{i}) \Theta (0)g(\lambda -z_{i}-\eta
)\right] \prod_{\alpha =1}^{n}\alpha(\lambda_{\alpha },\lambda ) \quad \quad
\label{eq2.28}
\end{eqnarray}
\[
\Lambda _{1,l }^{\alpha}(\lambda |z;\lambda _{1},\cdots ,\lambda _{n})
=\beta _{l+1}(\lambda ,\lambda _{\alpha })\prod_{i=1}^{N} \left[\rho
(\lambda _{\alpha}-z_{i}) \Theta (0)g(\lambda _{\alpha}-z_{i}+\eta )\right]
\prod\sb{\beta =1 \atop \beta \neq \alpha} ^{n} \alpha(\lambda _{\alpha
},\lambda _{\beta }) 
\]
\[
\Lambda _{2,l }^{\alpha}(\lambda |z;\lambda _{1},\cdots ,\lambda _{n})
=-\beta _{l-1}(\lambda ,\lambda _{\alpha }) \prod_{i=1}^{N}\left[\rho
(\lambda _{\alpha}-z_{i})\Theta (0) g(\lambda _{\alpha}-z_{i}-\eta )\right]
\prod\sb {\QATOP{\beta =1 }{\beta \neq \alpha}} ^{n} \alpha(\lambda _{\beta
},\lambda _{\alpha }) 
\]
and 
\begin{equation}
\Psi _{l}^{(\alpha )}=\Psi _{l}(\lambda _{1},\cdots \lambda _{\alpha-1},
\lambda , \lambda _{\alpha +1},\cdots \lambda _{n})  \label{eq2.29}
\end{equation}
The equation (\ref{eq2.27}) will be referred as OSBAE for 8-vertex model. It
is easy to show that $\sum_{l=-\infty }^{\infty }\omega ^{l}\Psi _{l}$ is an
eigenstate of the $T$ with eigenvalue 
\begin{equation}
\Lambda =\omega \Lambda _{1}(\lambda |z;\lambda _{1},\cdots ,\lambda
_{n})+\omega ^{-1}\Lambda _{2}(\lambda |z;\lambda _{1},\cdots ,\lambda _{n})
\label{eq2.30}
\end{equation}
provided that the $\lambda _{j}$ satisfy the Bethe ansatz equations 
\begin{equation}
\prod_{i=1}^{N}\frac{g(\lambda _{\alpha}-z_{i}+\eta )}{g(\lambda
_{\alpha}-z_{i}-\eta )}=\omega ^{-2}\prod\sb {\QATOP{\beta =1 }{\beta \neq
\alpha}} ^{n}\frac{\alpha (\lambda _{\beta }, \lambda _{\alpha})}{\alpha
(\lambda_{\alpha},\lambda _{\beta })}  \label{eq2.31}
\end{equation}
where $\omega =\exp (2\pi i\theta )$, with $0\leq \theta \leq 1$. But, in
what follows, we'll not consider the diagonalization problem of transfer
matrix $T$ and therefore we'll not impose Bethe ansatz equations (\ref
{eq2.31}) on the parameters $\lambda_{\alpha}$. We'll use the quasi
classical limit $\eta \rightarrow 0$ of (\ref{eq2.27}) to solve some system
of linear differential equation, which is an elliptic generalization of
Knizhnik-Zamolodchikov equation.

\section{Semiclassical limit of the OSBAE and the XYZ Gaudin magnets}

By semiclassical limit one understands the expansion of the vertex weight $%
{\cal R}(\lambda ,\eta )$ around the point $\eta _{0}$, such that ${\cal R}
(\lambda ,\eta _{0})=I\otimes I$ \cite{BD}. One can parameterize $\eta $,
such that $\eta _{0}=0$. In this subsection we examine the asymptotic
behavior of the OSBAE (\ref{eq2.27}) when $\eta \rightarrow 0$. >From the
equation (\ref{eq2.4}) we obtain the asymptotic behavior of the coefficients 
$w_{i}$: 
\begin{eqnarray}
w_{1}(\lambda -z_{i})=\eta J_1^i (\lambda)+{\rm o}(\eta^{2})  \nonumber \\
w_{2}(\lambda -z_{i})=\eta J_2^i (\lambda)+{\rm o}(\eta^{2})  \nonumber \\
w_{3}(\lambda -z_{i})=\eta J_3^i (\lambda)+{\rm o}(\eta^{2}) ,  \label{eq3.1}
\end{eqnarray}
where 
\begin{eqnarray}
\quad J_{i}^{1} &=& \frac{1+k{\rm sn}^{2}(\lambda -z_{i})} {{\rm sn}(\lambda
-z_{i})} ,  \nonumber \\
\quad J_{i}^{2} &=& \frac{1- k{\rm sn}^{2}(\lambda -z_{i})} {{\rm sn}%
(\lambda -z_{i})} ,  \nonumber \\
\quad J_{i}^{3} &=&\ \frac{{\rm cn}(\lambda -z_{i})\ {\rm dn} (\lambda
-z_{i})}{{\rm sn}(\lambda -z_{i})} .  \label{eq3.2}
\end{eqnarray}
Here we have used the following theta-function relations 
\begin{equation}
\begin{array}{lll}
\frac{d}{d\lambda }{\rm sn}\lambda ={\rm cn}\lambda {\rm dn}\lambda &  & 
\frac{d}{d\lambda }{\rm cn}\lambda =-{\rm sn}\lambda {\rm dn}\lambda \\ 
&  &  \\ 
\frac{d}{d\lambda }{\rm dn}\lambda =-k^{2}{\rm sn}\lambda {\rm cn}\lambda & 
&  \\ 
&  &  \\ 
{\rm sn}^{2}\lambda +{\rm cn}^{2}\lambda =1 &  & k^{2}{\rm sn}^{2}\lambda + 
{\rm dn}^{2}\lambda =1
\end{array}
\label{eq3.3}
\end{equation}
Simple calculations give us the following $\eta $-expansion for the transfer
matrix (\ref{eq2.11}) 
\begin{eqnarray}
T_N (\lambda |z)=A_{N}(\lambda |z)+D_{N}(\lambda |z)\equiv 2+2\eta^2
T_N^{(2)}+{\rm o}(\eta ^{3}) ,  \nonumber \\
T_N^{(2)}=\sum_{i<j} \left\{ J_{i}^{1}J_{j}^{1}\ \sigma _{i}^{1}
\sigma_{j}^{1}+J_{i}^{2}J_{j}^{2} \sigma _{i}^{2}\sigma_{j}^{2}+
J_{i}^{3}J_{j}^{3} \sigma _{i}^{3}\sigma _{j}^{3}\right\} .  \label{eq3.4}
\end{eqnarray}
As usual $T_N^{(2)}(\lambda |z)$ can be viewed as generating (operator)
function for the Gaudin magnet hamiltonians $H_i$ \cite{Gaudin} 
\begin{eqnarray}
H_i=res_{\lambda \rightarrow z_i}T_N=\sum_{j\neq i}\left\{ J_{ij}^1\sigma
_{i}^{1}\otimes \sigma _{j}^{1}+J_{ij}^2 \sigma _{i}^{2}\otimes \sigma
_{j}^{2}+J_{ij}^3 \sigma _{i}^{3}\otimes \sigma _{j}^{3}\right\} ,
\label{eq3.5}
\end{eqnarray}
where $J_{ij}^a=J_i^a (z_i) $. Here we observe that in the limit $%
k\rightarrow 0$, these expressions degenerate to the trigonometric
expressions presented in reference\cite{B2}. To find quasi classical version
of OSBAE (\ref{eq2.27}) we need also consider the quasi classical limit of
the $B_{k,l}(\lambda |z;s,t)$ operators and the global vacuum states $\Omega
_{N}^{l}$. From (\ref{eq2.19}) using (\ref{eq3.1}), (\ref{eq3.2}) we get the
following semiclassical expansion for the $B_{k,l}(\lambda |z;s,t)$
operators and for the semiclassical vacuum states 
\[
B_{k,l}(\lambda |z;s,t)=\eta^n B_{k,l}^{(1)}(\lambda |z;s,t)+ {\rm o}(\eta
^{n+1}) , 
\]

\[
B_{k,l}^{(1)}(\lambda |z;s,t)= \frac{g^{\prime}(0)g(t-\lambda -K)}{%
g^{2}(w)g(u+\lambda)}(l-k)+ 
\]
\begin{eqnarray}
+\frac{g(K)}{2g(w)g(u+\lambda)}\sum_{i=1}^{N}\left\{ [H^{2}( t-\lambda
)-\Theta^{2}(t-\lambda )]J_{i}^{1}\ \sigma_{i}^{1}+\right.  \nonumber \\
\left. +i[H ^{2}(t-\lambda )+\Theta^{2}(t-\lambda)]J_{i}^{2}\ \sigma
_{i}^{2}- 2H(t-\lambda) \Theta (t-\lambda)J_{i}^{2} \sigma _{i}^{3}\right\}
\label{eq3.6}
\end{eqnarray}
\[
\Omega _{N}^{(l)} =\Omega _{N}+{\rm o}(\eta ) , 
\]
\begin{equation}
\Omega _{N}=\otimes _{i=1}^{N}\left\{ H(s+z_{i}){\rm e}_{i}^{+}+
\Theta(s+z_{i}){\rm e}_{i}^{-}\right\}  \label{eq3.7}
\end{equation}
Taking the residues in the pole $\lambda =z_{i}$ of (\ref{eq2.27}), we get 
\begin{equation}
H_{i}\Phi =(\partial_w +h_{i})\Phi -2\sum_{\alpha=1}^{n}
(\partial_w+f_{\alpha})\Phi_{\alpha,i} ,  \label{eq3.8}
\end{equation}
where 
\begin{eqnarray}
h_{i}=\frac{1}{2}\sum_{j\neq i}^{N}log^{\prime }g(z_{i}-z_{j}) -\sum_{\alpha
=1}^{n}log^{\prime }g(z_i-\lambda _{\alpha })  \label{eq3.9} \\
f_{\alpha }=-\sum_{\beta \neq \alpha }^{n}log^{\prime }g(\lambda _{\alpha
}-\lambda _{\beta })+ \frac{1}{2}\sum_{j=1}^{N}log^{\prime }g(\lambda
_{\alpha }-z_{j}) ,  \label{eq3.10}
\end{eqnarray}
\begin{equation}
\Phi=B_{l-1,l+1}^{(1)}(\lambda _{1})\cdots
B_{l-n,l+n}^{(1)}(\lambda_{n})\Omega  \label{eq3.11}
\end{equation}
\[
\Phi_{\alpha,i}=B_{l-1,l+1}^{(1)}(\lambda _{1})\cdots
B_{l+1-\alpha,l-1+\alpha}^{(1)}(\lambda_{\alpha-1}) B_{\alpha,i} \times 
\]
\begin{equation}
\times B_{l-1-\alpha,l+1+\alpha}^{(1)}(\lambda_{\alpha+1}) \cdots
B_{l-n,l+n}^{(1)}(\lambda_{n})\Omega ,  \label{eq3.12}
\end{equation}
\begin{equation}
B_{\alpha,i}=res_{\lambda=z_i}B_{l-\alpha,l+\alpha}^{(1)} \frac{%
g^{\prime}(0)g(z_i-\lambda_{\alpha}+w)}{g(w)g(\lambda_{\alpha}-z_i)}.
\label{eq3.13}
\end{equation}
In (\ref{eq3.8}) have used the relation 
\begin{equation}
\Psi_{l \pm 1 }(\lambda|z;s,t)-\Psi_l (\lambda |z;s,t)=\pm 2\eta^{n+1}
\partial_w \Phi (\lambda |z;s,t)+{\rm o}(\eta^{n+2}) ,  \label{eq3.14}
\end{equation}
which is a direct consequence of the remarkable fact, that 
\begin{equation}
\Psi_{l \pm 1 }(\lambda|z;s,t)=\Psi_l (\lambda |z;s\pm 2\eta,t\pm 2\eta) .
\label{eq3.15}
\end{equation}

One may use the equations (\ref{eq3.8})-(\ref{eq3.13}) to obtain the
algebraic Bethe ansatz solution for the spin-$1/2$ XYZ Gaudin model. Indeed,
using the periodicity of $\Phi$, $\Phi_{\alpha ,i}$ over the parameter $w$
with period $4K$, it is easy to see, that Fourier component $\Phi_{l}$ of
the semi classical Bethe wave vector $\Phi$ 
\begin{equation}
\Phi _l = \int_{-2K}^{2K} e^{\frac{i\pi l w}{2K}}\Phi dw  \label{eq3.16}
\end{equation}
is the eigenvector of Gaudin hamiltonian $H_i$ with eigenvalue $h_i +i\pi
l/2K$, provided that the Bethe anzatz equation 
\begin{equation}
f_{\alpha}+i\frac{\pi l}{2K}=0  \label{eq3.17}
\end{equation}
for the parameters $\lambda _{\alpha}$ are valid. But our purpose in this
article is to obtain solutions for an elliptic generalization of KZ
equation, and therefor in what follows, we'll use (\ref{eq3.8}) without
imposing the condition (\ref{eq3.17}).

\section{Elliptic Knizhnik-Zamolodchikov equation}

The off-shell Bethe ansatz equations corresponding to rational and
trigonometric Gaudin magnets provide solution for the following differential
equation \cite{B0}-\cite{B3}, \cite{Cherednik}, \cite{SchV} 
\begin{equation}
\kappa \frac{\partial \Psi}{\partial z_i}= \sum _{j\neq
i}^Nr_{i,j}(z_i-z_j)\Psi  \label{eq5.0}
\end{equation}
Where $r_{i,j}(\lambda)$ is trigonometric or rational solution of quasi
classical Yang-Baxter equation and operator $H_i=\sum _{j\neq i}^N
r_{i,j}(z_i-z_j)$ coincides with Gaudin magnet Hamiltonian. Recall, that
self consistency condition of (\ref{eq5.0}) is the quasi classical
Yang-Baxter equation for $r_{i,j}(\lambda)$. Of course, elliptic $%
r_{i,j}(\lambda)$ also obeys the quasi classical Yang-Baxter equation, but
corresponding differential equation (\ref{eq5.0}) can not be solved via
off-shell Bethe ansatz method. The main reason is the fact, that the quasi
vacuum $\Omega _N$ of the Gaudin elliptic magnet ( \ref{eq3.5}), essentially
depends on inhomogeneities $z_1,\cdots z_N$ and auxiliary parameter $s$ (see
(\ref{eq3.7})). The elliptic generalization of (\ref{eq5.0}) is considered
in \cite{Et} but till now, an integral representation of this differential
equation is not constructed. In this section via off-shell Bethe ansatz
method we are constructing the solution for the following modification of (%
\ref{eq5.0}) which we'll call the Elliptic Knizhnik-Zamolodchikov (EKZ)
equation: 
\begin{equation}
\kappa \nabla _j\Psi= H_{j}\Psi-\frac{\partial}{\partial w}(\Sigma_{j}\Psi),
\label{eq5.1}
\end{equation}
where 
\begin{equation}
\nabla_j=\frac{\partial}{\partial z_j}-\Gamma_j,  \label{eq5.2}
\end{equation}
The operators $\Gamma_j$ and $\Sigma_j$ act on two dimensional space $V_j$
and have the following matrix elements 
\begin{eqnarray}
\Gamma_{11}=\frac{1}{2}log^{\prime}g(w)+ \frac{1}{2}log^{\prime}g(z_j+u);
\Gamma_{12}=-\frac{g^{\prime}(0)g(w-u-z_j)}{2g(w)g(z_j+u)},  \nonumber \\
\Gamma_{21}=-\frac{g^{\prime}(0)g(w+u+z_j)}{2g(w)g(z_j+u)}, \Gamma_{22}=-%
\frac{1}{2}log^{\prime}g(w)+ \frac{1}{2}log^{\prime}g(z_j+u), \\
\Sigma_{11}=\frac{g(w+K)g(u+z_j-K)}{g(w)g(u+z_j)}; \Sigma_{12}=\frac{%
g(K)H(s+z_j)H(t-z_j)} {g(w)g(u+z_j)},  \nonumber \\
\Sigma_{21}=-\frac{g(K)\Theta(s+z_j)\Theta(t-z_j)} {g(w)g(u+z_j)};
\Sigma_{22}=-\frac{g(w+K)g(u+z_j-K)}{g(w)g(u+z_j)}.  \label{eq5.3}
\end{eqnarray}
The solution of the EKZ equation (\ref{eq5.1}) is given by the multiple
integral 
\begin{equation}
\Psi =\oint \cdots \oint {\cal X}(\lambda |z)\Phi (\lambda |z;s,t)d\lambda
_{1}\cdots d\lambda _{n},  \label{eq5.4}
\end{equation}
where $\Phi$ is the quasi classical Bethe wave vector (\ref{eq3.11}) and $%
{\cal X}(\lambda |z)$ is a solution of the following self consistent system
of differential equations 
\begin{equation}
\kappa\partial_{z_j} {\cal X}=h_{j} {\cal X},  \label{eq5.5}
\end{equation}
\begin{equation}
\kappa\partial_{\lambda_{\alpha}}{\cal X}=-2f_{\alpha} {\cal X}.
\label{eq5.6}
\end{equation}
It is easy to see that solution is given by 
\begin{equation}
{\cal X}(\lambda |z)=\prod_{j<i}^{N}g(z_{j}-z_{i})^{1/2\kappa }\prod_{\alpha
<\beta }^{n}g(\lambda _{\alpha }-\lambda _{\beta })^{2/\kappa
}\prod_{j,\gamma }g(z_{j}-\lambda _{\gamma })^{-1/\kappa },  \label{eq5.7}
\end{equation}
Integration in (\ref{eq5.4}) is over closed, homologically nontrivial
circles of the set ${\bf C}^n\setminus {\bf D} \subset {\bf C}^n$, where $%
(\lambda_1,\cdots,\lambda_n)\in {\bf D}$  if and only if $\lambda_\alpha=z_j$
or $\lambda_\alpha=\lambda_\beta$ or $\lambda_\alpha=u$ for some $%
\alpha,\beta\in\{1,2,\cdots,n\}$ and $j\in\{1,2,\cdots,N\}$. The proof, that 
$\Psi(z_1,\cdots,z_N|s,t)$ given by (\ref{eq5.4}) indeed is solution of the
EKZ equation (\ref{eq5.1}) is straightforward, if one takes into account
OSBA equation (\ref{eq3.8}) and the relations 
\begin{equation}
\partial_{z_j}\Phi=\Gamma_j\Phi-\sum_{\alpha=1}^{n}
\partial_{\lambda_{\alpha}}\Phi_{\alpha,j},  \label{eq5.8}
\end{equation}
\begin{equation}
\Sigma_j\Phi=\Phi-2\sum_{\alpha=1}^{n}\Phi_{\alpha,j}.  \label{eq5.9}
\end{equation}
The identities can be proved directly, but it is easier first to perform
gauge transformation $\Phi \to {\bf G}\Phi$, where 
\begin{equation}
{\bf G}={\bf G}_1\otimes \cdots {\bf G}_N
\end{equation}
\label{eq5.10} and 
\begin{equation}
{\bf G}_i=\left( 
\begin{array}{ll}
H(s-z_i) & H(t-z_i) \\ 
\Theta(s+z_i) & \Theta(t-z_i)
\end{array}
. \right)  \label{eq5.11}
\end{equation}
This will be done in the next section.

\section{Elliptic Knizhnik-Zamolodchikov equation and KZB equation}

In this section we are going to show, that under some appropriately chosen
gauge transformation Elliptic KZ equation transforms into KZB equation.
Recall that KZB equation is linear differential equation for N-point
correlation function of WZNW model on torus. $\Gamma_j$ does not depend on $%
z_i$, $i\neq j$, and acts non trivially only on $j$-th component of tensor
product, therefore the connection $\Gamma$ is flat 
\begin{equation}
\left[ \nabla_j, \nabla_i\right]=\frac{\partial}{\partial z_j}\Gamma_i- 
\frac{\partial}{\partial z_i}\Gamma_j-\left[\Gamma_j,\Gamma_i\right]= 0
\label{eq6.1}
\end{equation}
This means that there exists some gauge transformation $G(z|s,t)$ such that 
\begin{equation}
\Gamma _j^G={\bf G}^{-1}\Gamma_j {\bf G}+\left(\partial_j {\bf G}%
^{-1}\right) {\bf G}\equiv 0  \label{eq6.2}
\end{equation}
Now it is not difficult to see that ${\bf G(z|s,t)}$, which satisfies above
relation (\ref{eq6.2}) is given by (\ref{eq5.10}), (\ref{eq5.11}). In other
words we have 
\begin{equation}
{\bf G}^{-1}\nabla_j {\bf G}= \frac{\partial}{\partial z_j}  \label{eq6.3}
\end{equation}
To find gauge transformed version of EKZ, we also need expressions for ${\bf %
G}^{-1}H_j {\bf G}$ and ${\bf G}^{-1}\Sigma_j {\bf G} $. By direct
calculation we have 
\begin{eqnarray}
{\bf G}^{-1}H_j {\bf G}=\Omega_j+\left({\bf G}^{-1}\partial_w {\bf G}- \frac{%
N}{2}\log^{\prime}g(w)\right)\sigma_j^3 \\
{\bf G}^{-1}\Sigma_j {\bf G}=\sigma_j^3,  \label{eq6.4}
\end{eqnarray}
Where 
\begin{eqnarray}
\Omega_j=\sum_{k\neq j}^N \left[ \frac{1}{2}\log^{\prime}g(z_{jk})
\sigma_j^3\otimes \sigma^3_k+\right.  \nonumber \\
+\left. \frac{g^{\prime}(0)g(w+z_{jk})}{g(w)g(z_{jk})}\sigma_j^- \otimes
\sigma_k^+ +\frac{g^{\prime}(0)g(w-z_{jk})}{g(w)g(z_{jk})}
\sigma_j^+\otimes\sigma_k^- \right] .  \label{eq6.5}
\end{eqnarray}
Using (\ref{eq6.3})-(\ref{eq6.5}) one easily obtains that $\phi \equiv{\bf G}%
^{-1}\Psi$ obeys the following differential equation 
\begin{equation}
\kappa \frac{\partial}{\partial z_j}\phi = \Omega _j \phi - \sigma_j^3
\partial_w \phi  \label{eq6.6}
\end{equation}
which is nothing but the KZB equation in special case of $sl(2,c)$ algebra,
when all representations are taken fundamental. Now let us perform gauge
transformation on the solution of EKZ equation (\ref{eq5.4}). For ${\bf G}%
^{-1}B_{k,l}(\lambda){\bf G}$ one obtains 
\begin{eqnarray}
{\bf G}^{-1}B_{k,l}(\lambda){\bf G}=\frac{g^{\prime}(0)g(t-\lambda -K)} {%
g(w)g(u+\lambda)}\left[(l-k)-\sum_{i=1}^N \sigma_i^3\right]+  \nonumber \\
+\sum_{i=1}^N \frac{2g^{\prime}(0)g(w-\lambda + z_i)} {g(w)g(\lambda -z_i )}%
\sigma_i^- \equiv \beta_{l-k}(\lambda),  \label{eq6.7}
\end{eqnarray}
where $\sigma^{\pm}=(\sigma^1 \pm \sigma^2)/2$. Similarly acting on $\Omega$
by the operator ${\bf G}^{-1}$ we obtain 
\begin{equation}
{\bf G}^{-1}\Omega= e_1^+ \otimes e_2^+ \otimes \cdots \otimes e_N^+ \equiv
|0>.  \label{eq6.8}
\end{equation}
Combining (\ref{eq3.11}), (\ref{eq6.7}) and (\ref{eq6.8}) we get the
following expression for the gauge transformed Bethe wave vector 
\begin{eqnarray}
{\bf G^{-1}\Phi}=\beta_2 (\lambda_1)\beta_4 (\lambda_2)\cdots \beta_{2n}
(\lambda_n) |0>=  \nonumber \\
=\widehat{\beta}(\lambda_1)\widehat{\beta}(\lambda_2) \cdots \widehat{\beta}%
(\lambda_n)|0> ,  \label{eq6.9}
\end{eqnarray}
where we have introduced the notation $\widehat{\beta}(\lambda)$ for the
second term in (\ref{eq6.7}) 
\begin{equation}
\widehat{\beta}(\lambda)=\sum_{i=1}^N \frac{2g^{\prime}(0)g(w-\lambda + z_i)%
} {g(w)g(\lambda -z_i )}\sigma_i^- .  \label{eq6.10}
\end{equation}
The second equality in (\ref{eq6.9}) follows from the fact, that first term
in (\ref{eq6.7}) does not give any contribution. So, finally for the
solution of KZB equation (\ref{eq6.6}) we have 
\begin{equation}
\phi =\oint \cdots \oint {\cal X}(\lambda |z)\widehat{\beta}(\lambda_1)
\cdots \widehat{\beta}(\lambda_n)|0> d\lambda _{1}\cdots d\lambda _{n},
\label{eq6.11}
\end{equation}
which is in complete agreement with the solution of KZB given by Felder and
Varchenko \cite{FV}.

\section{Conclusion}

In this paper an elliptic generalization of the Knizhnik-Zamolodchikov
equation is constructed. The main constructing block in this differential
equation is the quasi classical limit of the Baxters 8-vertex R-matrix, or
in an other words the elliptic Gaudin Hamiltonian. An integral
representation for the solution of this differential equation is found,
using off-shell Bethe ansatz method. It is shown that above mentioned
differential equation is connected with KZB equation via a gauge
transformation. It is worth noting that gauge transformation is the quasi
classical analogue of Baxter's famous 8-vertex-RSOS correspondence.

{\bf \flushleft  Acknowledgments} In the course of this work we benefited
from useful discussion with A.A.Belavin, R. Flume and one of us (H.B.) with
A. P. Veselov, M. Karowski, M. Schmidt. It is HB's pleasure to thank the
Departamento de F\'{i}sica da Universidade Federal de S\~{a}o Carlos for
hospitality and to FAPESP, Funda\c{c}\~{a}o de Amparo a Pesquisa do Estado
de S\~{a}o Paulo, for financial support.

\end{document}